\documentclass[none,article,submit,pdftex,moreauthors]{Definitions/mdpi} 
\renewcommand{\linenumbers}{}
\usepackage{subcaption}
\usepackage{colortbl}
\AtBeginDocument{%
  }

\Title{Gaze-Hand Steering for Travel and Multitasking in Virtual Environments}

\TitleCitation{Title}

\Author{\orcidA{}Mona Zavichi $^{1}$, André Santos$^{1}$, \orcidB{}Catarina Moreira $^{2}$, \orcidC{}Anderson Maciel $^{1}$ and \orcidD{}Joaquim Jorge$^{1}$*}

\AuthorNames{Mona Zavichi, André Santos, Catarina Moreira, Anderson Maciel and Joaquim Jorge}

\AuthorCitation{Zavichi, M.; Santos, A.; Moreira C.; Maciel, A.; Jorge, J.}

\address{%
$^{1}$ \quad INESC-ID/Instituto Superior Técnico/Universidade de Lisboa;
andre.santos@tecnico.ulisboa.pt, mona.zavichi@tecnico.ulisboa.pt, anderson.maciel@tecnico.ulisboa.pt\\
$^{2}$ \quad University of Technology Sydney; catarina.pintomoreira@uts.edu.au}

\corres{Correspondence: jorgej@tecnico.ulisboa.pt}

\abstract{
As head-mounted displays (HMDs) with eye-tracking become increasingly accessible, the need for effective gaze-based interfaces in virtual reality (VR) grows. Traditional gaze- or hand-based navigation often limits user precision or impairs free viewing, making multitasking difficult. We present a gaze-hand steering technique that combines eye-tracking with hand-pointing: users steer only when gaze aligns with a hand-defined target, reducing unintended actions and enabling free look. Speed is controlled via either a joystick or a waist-level speed circle. We evaluated our method in a user study (N=20) across multitasking and single-task scenarios, comparing it to a similar technique. Results show that gaze-hand steering maintains performance and enhances user comfort and spatial awareness during multitasking. Our findings support the use of gaze-hand steering in gaze-dominant VR applications requiring precision and simultaneous interaction. Our method significantly improves VR navigation in gaze-dominant, multitasking-intensive applications, supporting immersion and efficient control.
}

\hreflink{https://doi.org}

\firstpage{1}
\pubvolume{0}
\issuenum{0}
\articlenumber{0}
\pubyear{2025}
\copyrightyear{2025}
\externaleditor{Not applicable}
\datereceived{---}
\dateaccepted{---}
\datepublished{---}

\begin{document}
\maketitle

\begin{figure*}[h!]
  \includegraphics[width=\textwidth]{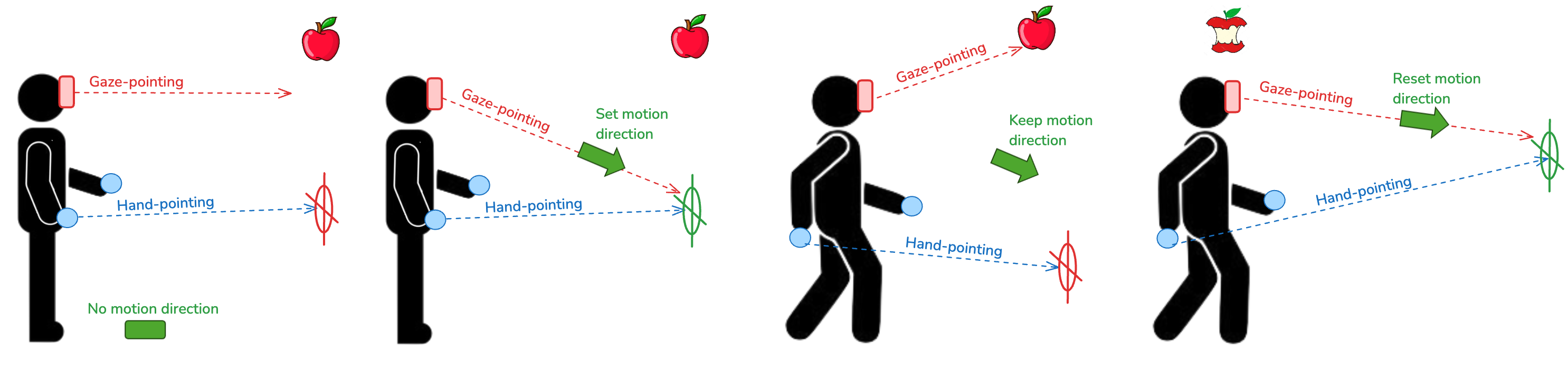}
  \caption{Travel direction is set only when gaze and hand pointing directions coincide. Travel continues in the direction set while hands and eyes can do other tasks, such as eating apples or popping balloons.}
  \label{fig:teaser}
\end{figure*}



\section{Introduction}
With affordable HMDs now incorporating eye tracking, effective gaze-dominant interfaces have become crucial for VR environments. Multitasking in VR poses challenges, as traditional gaze- or hand-based controls often compromise precision or limit the freedom to look around. We present a novel gaze-hand steering technique that addresses these issues with eye tracking and hand pointing.

In VR, navigation techniques are essential for effective user interaction as they enable the user to reach the location of objects that will be selected and manipulated within the virtual environment. Wayfinding is the element of navigation that helps users understand their position and plan their paths, often utilizing visual cues or maps. Travel, the motor aspect of navigation, uses specific techniques to allow users to determine the speed and direction with which their viewpoint will move through the virtual space. Although selection and manipulation tasks are frequently separated from navigation, the user concludes navigation before starting manipulation, it is not rare that the application demands selection and manipulation during locomotion, another word for travel. Selection, for example, enables users to choose objects or destinations using gaze, gestures, or controllers during motion. These elements enable users to explore and interact seamlessly within VR, enhancing immersion and spatial awareness.

Room size often limits the physical movement a user can make when applying locomotion in VR. Techniques such as joystick-based controls, gaze-directed movement, and walking-in-place have attempted to address these issues and enhance immersion without requiring large physical spaces. One relatively successful technique is the Magic Carpet~\cite{Medeiros2020}, which applies a flying carpet metaphor to separate direction and speed controls into two phases. It works with speed control methods such as a joystick, speed ring, and walking-in-place, combined with head-gaze as a pointing technique for direction control. This accommodates the naturalness of having a ground reference and the expeditiousness of flying. However, techniques relying solely on gaze direction for control have limitations. While gaze-oriented movement can provide precise and effortless steering~\cite{laviola2017}, it can also restrict users' ability to explore the environment freely, as gaze-based navigation often leads to unintentional actions, a phenomenon known as the "Midas Touch" problem~\cite{JacobMidas1990}. This limitation prevents users from looking around without unintentionally changing direction, impacting their overall navigation freedom.

On the other hand, while hand-pointing techniques offer precision, they prevent users from simultaneously using their hands for secondary tasks like selection or manipulation, which reduces their practicality in multitasking scenarios. Adding a trigger to switch between navigation-pointing and selection- or manipulation-pointing can make interactions cumbersome and increase the likelihood of errors.

In response to these limitations, our study proposes a technique combining eye-gaze and hand-pointing to enhance speed control and steering precision while supporting multitasking and offering a free look capability. This combined technique builds upon the strengths of gaze and hand-directed methods~\cite{Kang2024,Chen2023,Wagner2023,LystbText2022,LystbMenus2022}, enabling more seamless navigation that retains performance even during complex or multitasking activities.

Hence, we investigate the hypothesis that combining the gaze control techniques with pointing techniques can significantly improve speed and steering control. This potentially leads to the performance not being affected when exploring and performing other tasks while navigating, and leads to more accurate travel in the virtual environment, as gaze-oriented steering is considered to be the most efficient when more complex movements are involved~\cite{suma2007comparison,suma2009real}. 

We thus focus on answering the following research question: \textbf{How does combining gaze-directed steering with hand-pointing techniques allow exploration and multitasking without compromising performance?}

In the quest for responses, we conducted an experimental study with users to assess the user experience and travel efficiency achieved with the proposed technique. We also compare the performance of our method with two different speed control techniques—joystick and speed circle—and introduce new tasks to explore the free-look capability and selection interactions, supporting multitasking.

\section{Related Work}
\label{sec:relatedWork}

Travel is a critical task in VR, encompassing directional control and movement through virtual spaces. Steering-based methods are commonly used for VR navigation, where users can control the direction of movement as if steering a vehicle. However, traditional approaches that rely solely on gaze or hand input often suffer from limitations, such as physical fatigue or imprecise control, which recent studies aim to address by integrating gaze for quick directional indications and hand movements for refined speed and directional adjustments~\cite{Kang2024}. Despite these innovations, there is still a need for techniques that allow users to navigate while multitasking without compromising control precision or causing physical discomfort. Previous works have focused primarily on either selection or navigation but not adequately on how users can multitask during flight locomotion. 

\citet{Medeiros2020} studied flight locomotion by dividing the process into direction indication and speed control, utilizing a floor proxy with full-body representation. In the direction study, three techniques—Elevator+Steering, Gaze-Oriented Steering, and Hand Steering—were evaluated with 18 participants. The participants provided feedback through questionnaires, assessing user preferences, comfort, embodiment, and immersion. The Elevator+Steering technique enabled horizontal navigation through gaze and vertical navigation via buttons. Gaze-oriented and Hand-oriented techniques offered direction control based on head rotation and dominant hand movement, respectively. For the Speed study, they explored three techniques: joystick, speed circle, and WIP. The joystick controlled speed with an analog stick. The Speed Circle approach used the body as an analog stick, derived from a virtual circle metaphor~\cite{McMahan2012, Dam2013}. The last technique, adapted from \citet{BRUNO20171}, used knee movement for a maximum velocity 
of 5 m/s. 

The experiment involved navigating through a city scene from Unity3D Asset Store, with rings indicating the direction of the movement. The first experiment tested steering control with speed controlled by a button, and the second focused on speed control using a hand technique for direction. The first experiment showed that gaze and hand-oriented steering techniques had close results, with gaze showing shorter path length and total time and the hand technique with fewer collisions. In the second experiment, only the hand technique was used for direction control due to limited visual exploration with the gaze method.

Lai et al. investigated gaze-directed steering on user comfort in virtual reality environments~\cite{Lai2020comfort}. Previous research has been inconclusive about which technique is best. Some found Gaze-directed better than Hand-directed~\cite{suma2009evaluation, suma2007comparison}, while others found the opposite~\cite{Bowman1997, christou2016navigation}. They compared gaze, hand, and torso-directed steering, finding that gaze-directed methods significantly reduced simulator sickness symptoms like nausea and dizziness. \citet{zeleznik2005} explored gaze in Virtual Reality interactions, testing if gaze-oriented steering provides benefits over existing techniques. They argued that hand input in VR interactions is often redundant with gaze information, potentially reducing arm fatigue and improving interactions. They focused on terrain navigation, identifying speed control with an analog joystick as a main problem, and offering fixed speeds. Another issue was the inability to orbit around a region of interest. They proposed using gaze to lock areas of interest and orbit using a joystick or tablet gestures.

\citet{Lai2021} replicated a dual-task methodology to compare steering-based techniques with target-based ones, finding steering techniques afford greater spatial awareness due to continuous motion. However, steering techniques may increase cybersickness \cite{Hadgood2017}. VR developers are recommended to provide Gaze-directed steering as an intermediate option between novice-friendly teleport and expert-friendly Hand-directed steering.

\citet{Zielasko2020} mentioned the advantages of seated VR for ergonomics and long-term usage. They found torso and gaze steering effective for direction while leaning caused fatigue and cybersickness. However, gaze/view-directed steering has issues inspecting the environment independently of movement direction, commonly called the "Midas Touch" problem~\cite{JacobMidas1990}, where gaze interactions trigger unintended actions. \citet{Tregillus2017} addressed this by introducing head-tilt motions for independent "free-look" control. 

Combining gaze with other input modalities, particularly hand gestures, has emerged as a solution to overcome the limitations of gaze-only interaction. Techniques that utilize gaze for initial selection and confirm actions with a secondary hand gesture have effectively reduced the "Midas Touch" effect by decoupling selection from confirmation inputs~\cite{LystbText2022, LystbMenus2022}. \citet{LystbText2022} introduced Gaze-Hand Alignment, where the alignment of gaze and hand inputs triggers selection, offering a seamless and efficient approach to interaction tasks like menu selection in AR. This technique leverages the natural coordination of gaze and hand, allowing users to pre-select a target with their gaze and confirm it by aligning their hand, thereby reducing physical strain and increasing interaction speed~\cite{Wagner2023}. Recent advancements, such as the study by~\citet{Chen2023}, extend this approach by combining gaze rays with controller rays, effectively disambiguating target selection to reduce selection time and enhance accuracy. Similarly, the study by~\citet{Kang2024} combines gaze for initial direction setting with hand-based adjustments, allowing users to control navigation speed and direction without unwanted activations. Also, HMDs can impact spatial awareness and task performance, especially when body representation and perspective are manipulated~\cite{Medeiros18}.

The integration of gaze and hand inputs has improved significantly over single-modality systems, providing more natural, efficient, and accurate interaction in VR environments. Despite these advancements, multitasking within navigation, particularly during flight locomotion, needs further development. Our approach extends gaze-hand-directed steering techniques explored in previous studies, e.g. \citeauthor{Chen2023}, \citeyear{Chen2023}; \citeauthor{Kang2024}, \citeyear{Kang2024}; \citeauthor{Medeiros2020}, \citeyear{Medeiros2020}, introducing a mechanism that supports flight locomotion and enables users to navigate efficiently without losing orientation while multitasking.
\nocite{Correia05}

\section{Our Gaze-hand Steering Mechanism}
\label{sec:approach}

Hand pointing is an affordable and handy solution to indicate travel direction. However, currently, many HMDs incorporate eye-tracking capability, and several eye-dominant techniques are prevailing. One piece of evidence is that industry standards are adopting eye-gaze as the default dominant paradigm, as is the case with the high-end consumer device Apple Vision Pro\footnote{https://www.apple.com/apple-vision-pro/}. A major constraint in eye-gaze-directed techniques, however, is the free-look or "Midas Touch" problem~\cite{JacobMidas1990,Tregillus2017}, which makes it difficult for users to look around without affecting movement direction. To address this, we implemented a mechanism that changes between free-look (exploring the environment with gaze without changing movement direction) and directed look (gaze changes movement direction) that allows multitasking while traveling. This mechanism combines eye-tracking gaze techniques and hand-steering, combining the two vectors to decide the direction vector. Inspired by the gaze as a helping hand study~\cite{zeleznik2005}, we combine eye-gaze and hand pointing to lock direction and allow users to explore while maintaining movement direction. This approach uses an invisible string metaphor, where the direction locks with the interception of the eye and hand vectors, enabling users to move while looking around.

\subsection{Pointing direction}
The direction is null at the start, and no travel is possible. The dominant hand trigger activates the display of a target object 1 m wide at a 5 m distance along the hand-pointing direction. Then, a ray is cast in the direction of the eye-gaze, and if this ray intercepts the hand-driven target object, a new direction is set for travel (see Fig.~\ref{fig:teaser}). The chosen direction is that of the eye gaze, providing a more egocentric definition of the direction to follow. Both eye-gaze and hands can do other tasks such as selecting or manipulating objects at any time. A new direction will only be set when the two directions align while the dominant hand trigger is pressed.

\subsection{Controlling speed}
\begin{figure}[b]
    \centering
    \begin{minipage}{0.42\linewidth}
        \centering
        \includegraphics[width=\linewidth]{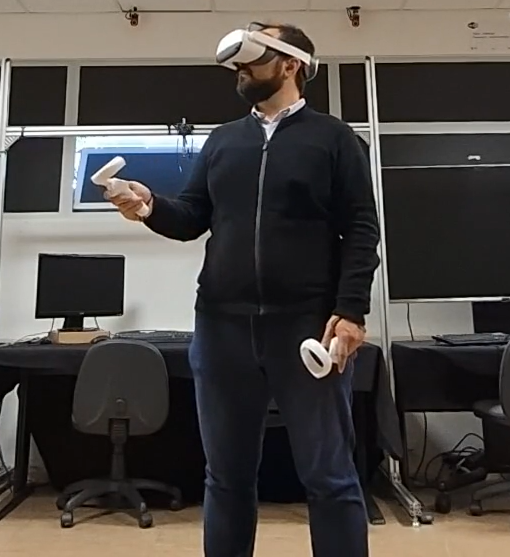}
        \caption{User using joystick}
        \label{fig:UserJoystick}
    \end{minipage}\hfill
    \begin{minipage}{0.42\linewidth}
        \centering
        \includegraphics[width=\linewidth]{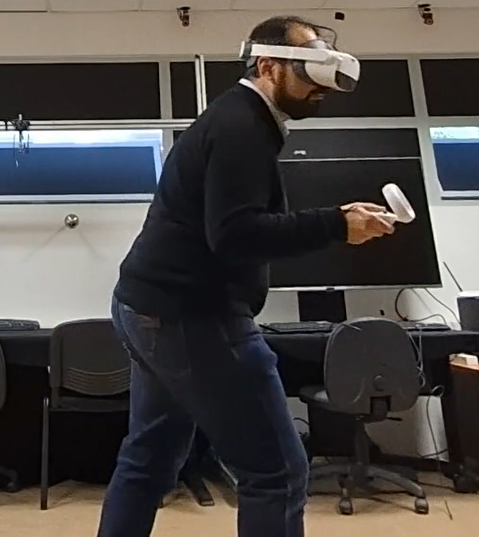}
        \caption{User using Speed Circle}
        \label{fig:UserSpeedCircle}
    \end{minipage}
\end{figure}

Whenever a direction is set, travel speed will determine the quantity of motion in that direction. The literature has plenty of solutions. We tested two simple techniques to control speed: the joystick and the speed circle. Joystick uses the analog thumbstick to smoothly set a velocity forward or backward. A full push on the stick will set the maximum speed, or 5 m/s, which is the same used by the Magic Carpet (see Figure~\ref{fig:UserJoystick}). Speed circle, in turn, uses the user's body as a joystick, i.e., leaning forward or backward sets a speed proportional to the amount of leaning (see Figure~\ref{fig:UserSpeedCircle}). Returning to the initial neutral position sets the speed to zero.

\subsection{Multitasking}
As we intend to use multitasking, the interface includes other simple actions. Eye gaze is the dominant input for selection. So, while it is used to select the hand-defined target for travel direction, it is also used to select other objects. One example is to point and pop balloons. This action is triggered when the eye gaze hits a balloon and the non-dominant hand trigger is pulled.  
To accurately shoot a target, users needed to look away from the directional scope and explore the environment to locate the targets. Upon identifying a target, they had to press the trigger on the controller in their non-dominant hand only when their line of sight intersected with the target object. This setup allowed users to multitask, managing navigation and target acquisition simultaneously.
\begin{figure}[t]
    \centering
    \begin{minipage}{0.42\linewidth}
        \centering
        \includegraphics[width=0.8\linewidth]{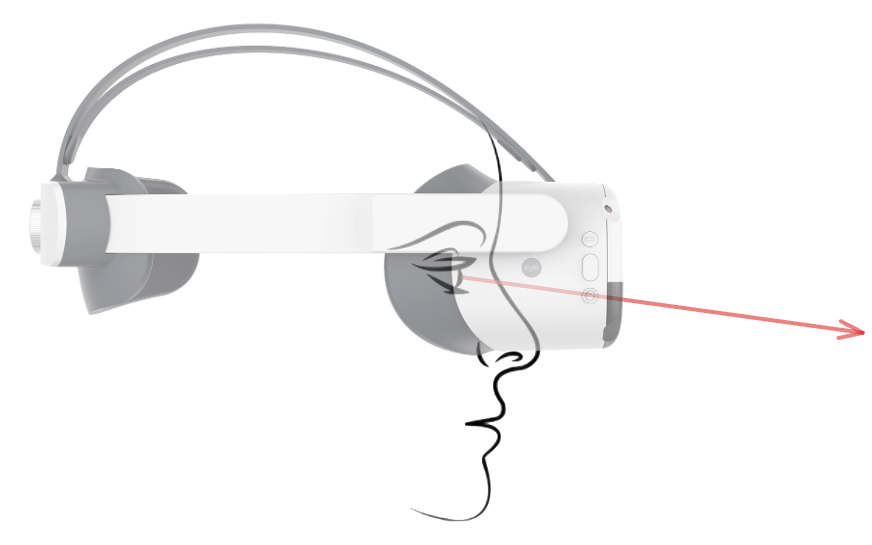}
        \caption{Eye gaze points objects for selection.}
        \label{fig:controllersE}
    \end{minipage}\hfill
    \begin{minipage}{0.48\linewidth}
        \centering
        \includegraphics[width=0.8\linewidth]{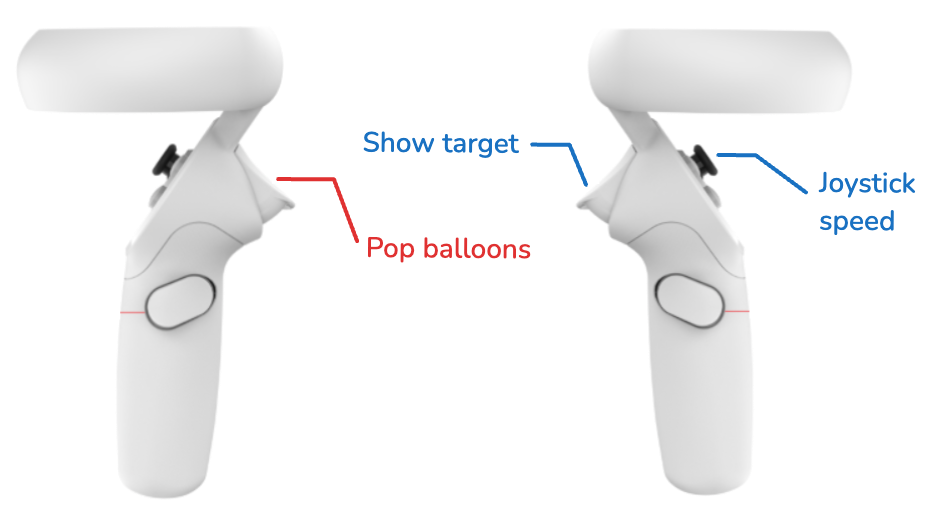}
        \caption{Left trigger confirms selection. Right trigger displays the target for travel.}
        \label{fig:controllersD}
    \end{minipage}
\end{figure}
Figures~\ref{fig:controllersE} and ~\ref{fig:controllersD} summarize the controllers and actions.

\subsection{Implementation}

We implemented the above techniques for the PICO Neo 3 Pro EYE VR Headset, which supports eye-tracking, and the two included hand controllers. The eye-tracking API used with that device is the Tobii Extended Reality toolkit. Nevertheless, any similar contemporary VR kit, such as the MetaQuest Pro, could alternatively be used.

We used Unity version 2021.3 for the virtual environment and retained most assets and design elements from Magic Carpet~\cite{Medeiros2020} to enable direct comparative evaluation. One difference is that we did not include the walking-in-place technique due to subpar performance results, and thus did not use the trackers they used. Another is that the Magic Carpet was based on a CAVE, whereas we used an HMD. Therefore, we also refined the Speed Circle technique to HMD environments. Our redesign sought to minimize user movement and reduce confusion regarding the circle’s center. To this end, the speed circle was positioned at waist level and resized to dimensions of 0.6 x 0.6 meters, aligning it with the user’s physical space and preventing abrupt accelerations during sharp turns. This enhanced stability ensures effectiveness and averts unintended movements by establishing a maximum speed threshold of 5 m/s and improving access to a null velocity zone. We also added graphical elements, such as redesigned semicircle colors with positive and negative symbols in each sector to indicate the areas for increasing or decreasing velocity. Leaning or moving towards the designated green area, marked with a plus sign, increases velocity, while moving towards the red semicircle, marked with a negative sign, decreases velocity.

As for the experimental tasks, which will be further discussed in Sec.~\ref{sec:eval}, we adjusted visibility to keep all rings visible, minimizing wayfinding difficulties to focus on the eye-tracking performance.

\section{Evaluation}
\label{sec:eval}

We designed and conducted a user study to assess and characterize the effectiveness and efficiency of our steering locomotion method. We seek to establish the feasibility of using eye-tracking steering and hand-pointing gestures for navigation, enabling users to freely explore the environment. Later, we also compare our results with those from the Magic Carpet technique~\cite{Medeiros2020}.

We anticipate observing effects in several interaction metrics. So, we measured total completion times for tasks, the occurrence of collisions, path lengths traveled, scores in the presence and workload assessments, and cybersickness. Furthermore, we anticipate introducing free look capabilities, facilitated by eye-tracking technology, that should not adversely impact flying or completion times. We expect flying and idle times to remain consistent across tasks, indicating that incorporating free-look functionality does not hinder overall performance. Additionally, we anticipate that users will perform simultaneous tasks involving traversing rings and destroying targets without undermining performance and experience when compared to data in the literature.

\subsection{Environment}

\begin{figure}[t]
    \centering
    \begin{minipage}{0.47\linewidth}
        \centering
        \includegraphics[width=\linewidth]{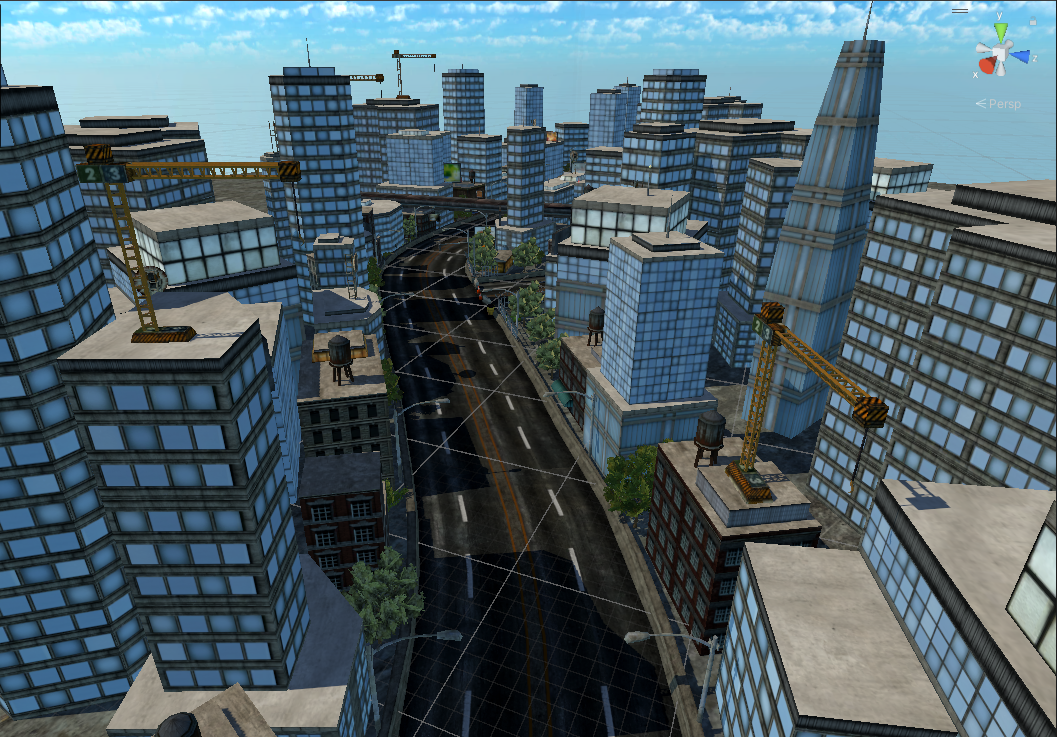}
        \caption{Perspective of the virtual environment.}
        \label{fig:cityScene}
    \end{minipage}\hfill
    \begin{minipage}{0.47\linewidth}
        \centering
        \includegraphics[width=\linewidth]{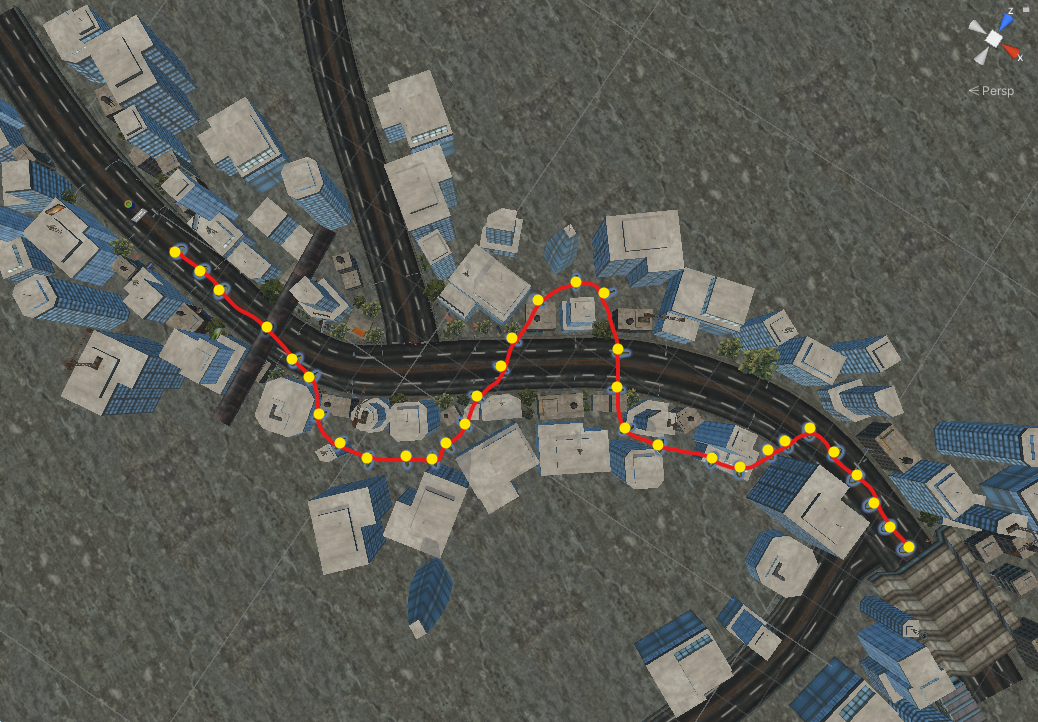}
        \caption{Path used in the experiment (350 m). Yellow dots indicate the rings placement.}
        \label{fig:cityAbove}
    \end{minipage}
\end{figure}

The virtual environment for this study is the same as that of the Magic Carpet study to facilitate direct comparison. The environment was based on a city scene with a path of rings through which the user must fly, as seen in Figs.~\ref{fig:cityScene} and~\ref{fig:cityAbove}, obtained from the Unity3D Asset Store. This scene was modified to remove visual clutter and present a smoother path for testing. The interaction was implemented with the help of Unity XR's interaction kit, a framework for implementing virtual and augmented reality functionality in Unity projects.

\subsection{Conditions}
We designed the experiment to evaluate the impact of gaze-hand steering in immersive environments by assigning participants specific tasks under varying conditions—each task condition aimed at different aspects of user interaction to assess steering and targeting capabilities more effectively. The following subsections describe the functions that participants performed during the study.
\subsubsection{Tasks}

There are two task conditions. One is a \textbf{single-task} race where participants fly through \textbf{Rings} scattered across the map, with obstacles between the rings to test effective steering without colliding. Rings make a sound when crossed to indicate the path. 

The other task condition is \textbf{multitask}. It involves a \textbf{Target} practice challenge in which users navigate through rings and shoot balloon targets placed at several locations along the path. Balloons are popped by looking at them and pressing the non-dominant hand trigger. This design focused on assessing the effectiveness of gaze-hand steering in isolation, providing insights into its performance characteristics. Comparisons with traditional eye-gaze steering are discussed using published data from the Magic Carpet experiment~\cite{Medeiros2020}.

\subsubsection{Interface conditions}

We implemented our gaze-hand steering as the only interface for travel direction. However, we use two conditions for speed control. Gaze-hand steering is coupled with one of the speed control interactions, making two interface conditions: Gaze-hand steering + joystick and Gaze-hand steering + speed circle. 

\subsubsection{Experimental conditions}

Our user study thus exposed participants to four different conditions:

\begin{itemize}
    \item Gaze-hand steering -- Joystick -- Rings
    \item Gaze-hand steering -- Joystick -- Targets
    \item Gaze-hand steering -- Speed circle -- Rings
    \item Gaze-hand steering -- Speed circle -- Targets
\end{itemize}

The four conditions use Gaze-hand steering to control direction. Arguably, having only one direction control condition in the protocol limits the results to findings that characterize the technique in isolation. However, we did not include other direction techniques to maintain a manageable session length and minimize participant fatigue. It would also be interesting to see how participants perform if simple eye-gaze steering is used compared to how they perform with our technique. We compare the two methods in Sec.~\ref{sec:MCcompare} using published data from the Magic Carpet experiment~\cite{Medeiros2020}. 

\subsection{Apparatus}\label{sec:apparatus}

To control for extraneous factors and ensure a consistent environment, all trials were conducted in a virtual reality laboratory with controlled lighting and sound levels to minimize distractions. Participants were equipped with a Head Mount Display (HMD) and calibrated hand controllers at the beginning of each session. Calibration ensured precise eye tracking, and participants were given a short training session to familiarize themselves with the gaze-hand alignment technique and verify the alignment of the equipment.

We collected both objective and subjective measures. The system logged performance measures in four aspects: task speed (time to complete), collisions (with obstacles), path length, and movement time (percentage of time passed moving instead of idling). We also administered three utility and usability measurement scales: Slater-Usoh-Steed (SUS) presence questionnaire \cite{slater1994}, the Simulator Sickness Questionnaire (SSQ) \cite{kennedy1993simulator}, and the NASA Task Load Index (NASA-TLX) \cite{hancock1988human}. 

Some of these instruments were explicitly chosen to allow comparison with the results of the Magic Carpet study.

\subsection{Procedure}

The experimental procedure followed a within-subjects design, with participants experiencing the four conditions once. The interface condition presentation order was counterbalanced, and the task condition was fixed with the single-task Ring being performed first and the multitask Target being performed second.

After a brief explanation, participants signed an informed consent form and were told they could stop at any time if they felt discomfort, cybersickness, or other reasons. Then, they filled out a pre-test profile form. After that, they were introduced to the VR headset, the eye-tracking calibration was performed, and they were guided to the marked spot on the floor and allowed to perform the test task, where they were free to explore and practice. When they were comfortable, they started the experimental tasks.

After exposure to all four conditions, participants completed the post-test usability and experience questionnaires of Sect.~\ref{sec:apparatus} and a list of subjective questions about the system. We thanked the participants, and they were dismissed. 

The experience lasted approximately 40 minutes per participant, including the 15-minute briefing, training, and questionnaires. 

\subsection{Participants}

Twenty participants, aged 17 to 35 (\(\mu \approx 23.95\), \(\sigma \approx 5.36\)), volunteered for the study. The majority had at least a bachelor's degree. Despite most users having some VR experience, most were inexperienced, having used VR less than five times. Additionally, two-thirds of the participants never or sometimes experienced dizziness or nausea using VR, while the other third never used it.

\section{Results}

We present our results in this section according to the evidence found. We start by analyzing usability from subjective measurements and then go into the objective data collected. In the objective data part, we first look into our user performance data in isolation, developing along different tasks and interface conditions. Then, we compare our data with the results from the Magic Carpet~\cite{Medeiros2020} in Sec.~\ref{sec:MCcompare}.

\subsection{Subjective Scores}

This section evaluates the integrity of the developed system using SUS-presence, SSQ, and NASA-TLX questionnaires. It includes user impressions of each speed technique to compare against each other and understand how our steering approach performs overall. Statistical results are examined. Then, we present the users' responses to additional general questions regarding the experience.

A summary of the subjective instruments results from 20 participants is shown in Table~\ref{tab:questionnaire_metrics}, providing a Mean ($\mu$), Standard Deviation ($\sigma$), and 95\% Confidence Interval (CI). 

First, the average SUS-presence score is 15.54 out of 20 total points (77.7\%). Then, we observed that none of the participants felt severe sickness symptoms. Only four marked at least two of the 16 symptoms as moderate. The SSQ total score at the end of the participation is 21.25, but the nausea factor is lower, 13.87 on average. Disorientation was scored higher, at 30.37. For task load, the total average NASA-TLX score falls below 29 on a scale of 100, indicating a medium workload category.

\begin{table*}[h]
    \centering
    \caption{Summary of the subjective metrics}
    \label{tab:questionnaire_metrics}
    \begin{tabular}{l|l|c|c|c}
        \hline
        \textbf{Questionnaire} & \textbf{Metric} & \textbf{$\mu$} & \textbf{$\sigma$} & \textbf{CI} \\
        \hline
        SUS & Presence SUS Score & 15.54 & 3.55 & [14.06, 17.03] \\
        \hline
        SSQ & Cyber Sickness Score & 21.25 & 18.97 & [13.31, 29.18] \\
        \hline
         & Mental Demand & 2.41 & 1.05 & [1.94, 2.87] \\
         & Physical Demand & 1.55 & 0.73 & [1.22, 1.87] \\
        NASA-TLX & Temporal Demand & 2.91 & 1.34 & [2.31, 3.50] \\
         & Performance & 2.23 & 0.81 & [1.86, 2.58] \\
         & Effort & 1.54 & 0.73 & [1.22, 1.87] \\
         & Frustration & 1.45 & 0.73 & [1.12, 1.78] \\
         & TOTAL SCORE & 28.91 & 6.89 & [25.86, 31.96] \\
        \hline
    \end{tabular}
\end{table*}

\subsection{Additional subjective Impressions}

Table~\ref{tab:questionnaire_results} summarizes the participants' impressions of using the system on a five-point Likert scale. The responses indicate that walking inside the virtual circle is significantly easier (Wilcoxon = 28.5, \(p$=0.020$\)) with the joystick (avg$=4.36$, SD$=0.90$) than with the speed circle (avg$=3.55$, SD$=1.01$). Controlling speed was significantly easier (Wilcoxon$=32.0$, \(p$=0.033$\)) with the joystick (avg$=4.13$, SD$=0.83$) compared to the speed circle (avg$=3.31$, SD$=1.13$).

Moving around the virtual environment was also easier with the joystick (Wilcoxon = 12.5, \(p = 0.034\)). Users reported a higher sense of agency with the joystick (average = 4.32, SD = 3.45) compared to the speed circle (average = 3.45, SD = 1.14) (Wilcoxon = 0.0, \(p = 0.003\)). Users also felt safer inside the circle when using the joystick (Wilcoxon = 9.0, \(p = 0.046\)). These results collectively suggest that the joystick provides a more familiar user experience as is commonly used in video games.
However, some users preferred the speed circle for its more immersive experience, allowing direct interaction through body gestures. Notably, some users achieved better results with the speed circle. Despite this, the joystick was generally preferred for speed tasks. Some users found it easier to maintain movement with the speed circle during target practice, as the joystick required more frequent stops for slow movements. However, the speed circle was more prone to imbalance issues, requiring constant position adjustments when changing direction. One user commented, \textit{"I felt that with the speed circle I was more in control, but with the stick it was more immediate. In the speed circle, I was much more likely to overbalance and fall, whereas with the joystick I wasn't"}.

No significant differences were found between the two techniques regarding fatigue or fear of heights. Similarly, there were no significant differences in body ownership or sense of self-location. Both methods were perceived as equally effective in maintaining user immersion and safety.


\begin{table}[t]
   \centering
    \caption{Subjective impressions collected and presented as Median (Interquartile Range) Values}
    \label{tab:questionnaire_results}

\begin{tabular}{p{6.5cm}|c|c}
    \hline
    \textbf{Question} & \textbf{Joystick} & \textbf{Speed} \\
    \hline
    Q1. It was easy to walk inside the circle. & 5 (1) & 4 (1) \\
    Q2. It was easy to indicate the direction of movement. & 4.5 (1) & 4 (1.75) \\
    Q3. It was easy to control the speed of movement. & 4 (1) & 3.5 (1.75) \\
    Q4. It was easy to move around the VE. & 4 (1) & 4 (1) \\
    Q5. It was easy to reach the rings. & 4 (0.75) & 3.5 (1) \\
    Q6. It was easy to avoid obstacles. & 4 (1.75) & 3.5 (1.75) \\
    Q7. It was easy to coordinate movements. & 4 (1.75) & 3 (1) \\
    Q8. I felt safe inside the circle. & 5 (1) & 4 (1) \\
    Q9. I felt a fear of heights. & 0 (0.75) & 0 (1) \\
    Q10. I felt fatigue. & 0 (0) & 0 (0) \\
    Q11. I felt a sense of agency. & 4 (1) & 4 (1) \\
    Q12. I felt body ownership. & 4 (1) & 4 (1.75) \\
    Q13. I felt a sense of self-location. & 4 (1) & 4 (1) \\
    \hline
\end{tabular}

\end{table}

\subsection{Objective measurements}

This section examines task performance across different techniques and tasks. The analysis includes statistics with a 95\% confidence interval for each metric to compare techniques, tasks, and previous findings. 
Both objective metrics and subjective scores were analyzed to evaluate differences between techniques. The normality of the samples was assessed using the Shapiro-Wilk test. Although most conditions followed a normal distribution, the Targets\_SpeedCircle condition significantly deviated from normality ($p = 0.001$). To ensure consistency and robustness, we proceeded with the Kruskal–Wallis nonparametric test for all comparisons to identify significant effects.  Post hoc pairwise comparisons were performed using Dunn's test with Bonferroni correction to determine specific differences between conditions.
The main goal was to observe to what extent our technique supports a secondary task without penalizing performance outcomes.

\subsubsection*{\textbf{Time}}

Task completion time shows that the Joystick method is faster in both the Ring task ($t(38) = -4.172, p $<$ 0.01$) and the Target practice ($t(38) = -2.314, p = 0.026$). No significant differences were found between tasks using the Joystick ($t(38) = -0.179, p = 0.858$), but Speed Circle showed slightly better times in the Target practice ($t(38)=2.596, p=0.013$).

\begin{figure}[!b]
    \centering
    \includegraphics[width=0.8\linewidth]{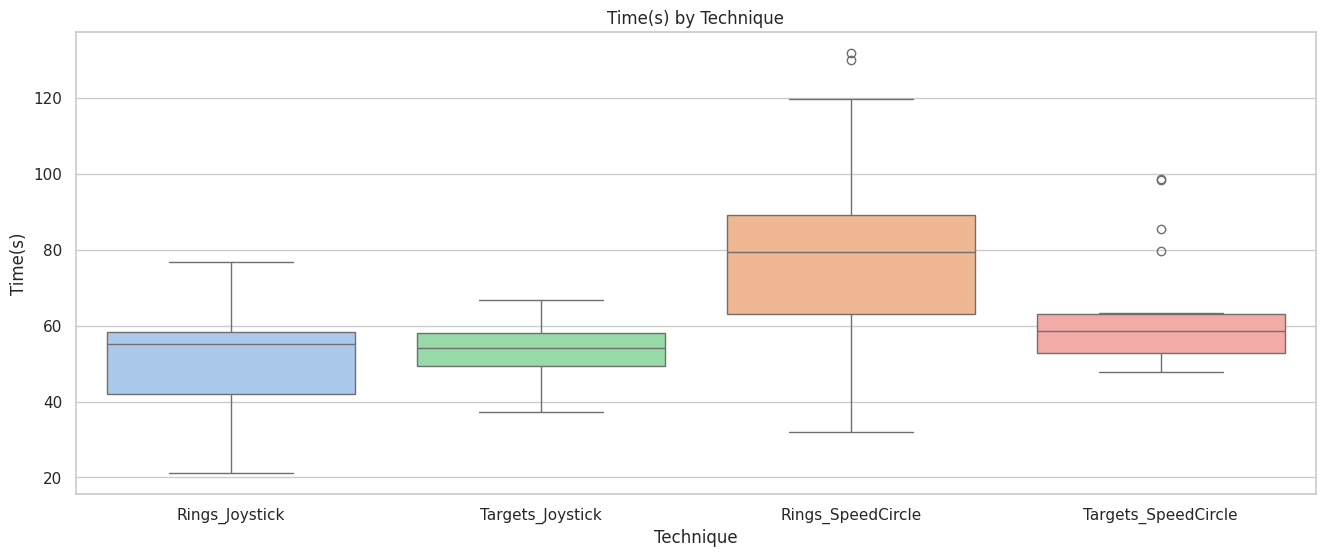}
    \caption{Time to complete each task (Ring and Target) with each speed technique (Joystick and Speed Circle).}
    \label{fig:BoxPlotTimes}
\end{figure}

Fig.~\ref{fig:BoxPlotTimes} shows that for the Target+Joystick task, the median completion time is approximately 55 seconds, with a mean slightly above the median, indicating a right-skewed distribution. The IQR is relatively narrow, suggesting consistent performance among participants. In the Target+Speed Circle task, the median completion time is around 60 seconds, with a mean close to the median, indicating a relatively symmetric distribution and consistent performance, with a few outliers. The Rings+Joystick task has a median completion time of about 50 seconds, with a mean close to the median. The IQR is narrow and shows consistent performance, although a few outliers indicate some variability. Rings+Speed Circle has the highest median completion time at around 80 seconds, with a mean slightly higher than the median, indicating a right-skewed distribution. The IQR is the widest among the conditions, suggesting significant variability in performance, with several outliers showing that some participants took much longer to complete the task.

\begin{figure*}[t]

    \begin{subfigure}[t]{.22\linewidth}
      \includegraphics[width=0.98\linewidth]{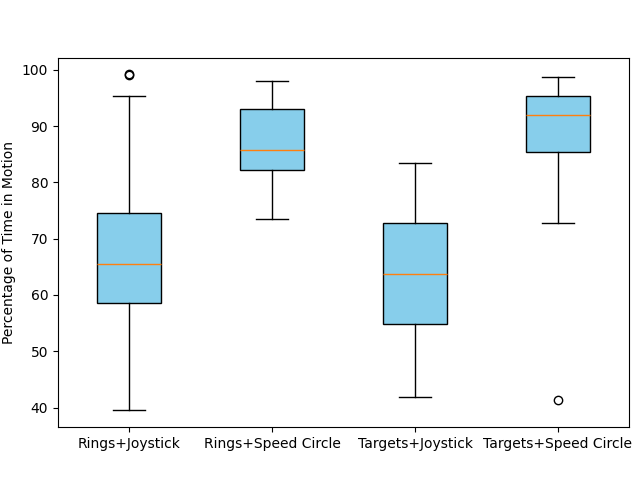}
        \caption{Flying vs. idle}
        \label{fig:flyingBoxPlot}
    \end{subfigure}
    \hfill
    \begin{subfigure}[t]{.22\linewidth}
      \includegraphics[width=0.98\linewidth]{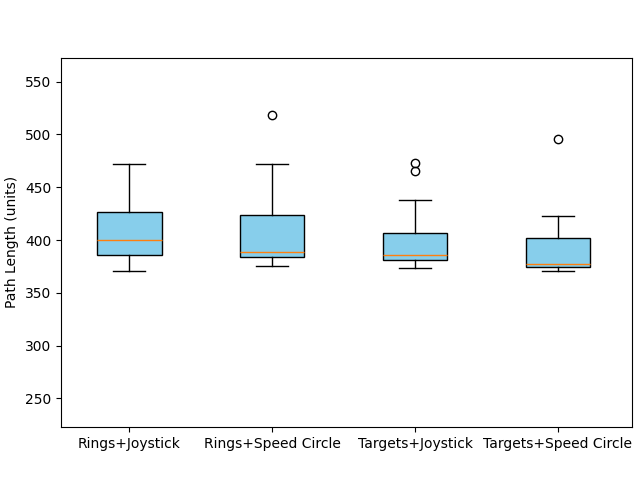}
        \caption{Path Length}
        \label{fig:pathBoxPlot}
    \end{subfigure}
    \hfill
    \begin{subfigure}[t]{.22\linewidth}
      \includegraphics[width=0.98\linewidth]{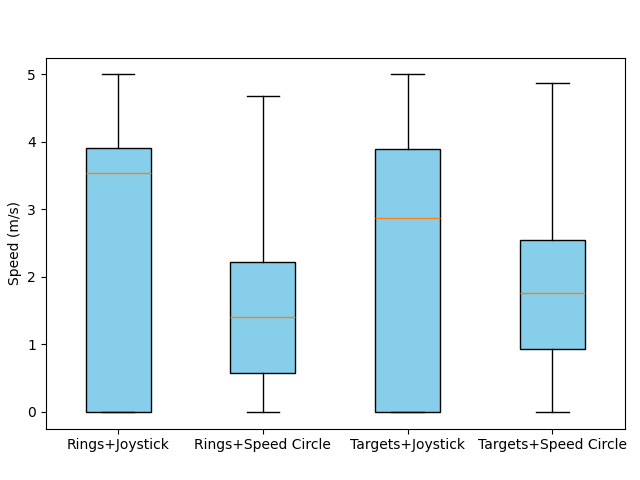}
        \caption{Overall Speed}
        \label{fig:speedBoxPlot}
    \end{subfigure}
    \hfill
    \begin{subfigure}[t]{.28\linewidth}
      \includegraphics[width=0.98\linewidth]{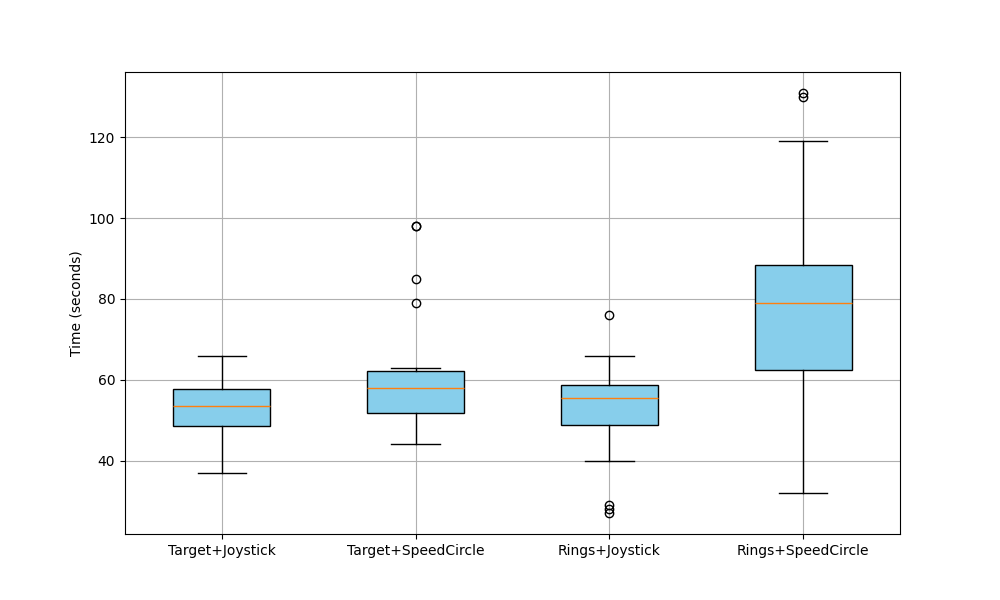}
        \caption{Collisions}
        \label{fig:collisionsBoxPlot}
    \end{subfigure}



   \caption{Comparison of performance metrics across joystick and speed circle techniques, including ring navigation and target shooting tasks, to evaluate multitasking performance during VR navigation: (a) percentage of time in motion (Flying vs. Idle), (b) total path length (units), (c) average navigation speed (m/s), and (d) total collision time (seconds).}
   \label{fig:boxplotmetrics}
\end{figure*}

\subsubsection*{\textbf{Flying vs. Idling}}
See Fig.~\ref{fig:flyingBoxPlot}. Speed circle tasks had higher flying percentages for both Ring (87.072, t(38) = -4.921, p $<$ 0.001) and Target tasks (87.928, t(38) = -6.259, p $<$ 0.001) than the joystick method. Idle time was significantly higher with the joystick, indicating more consistent movement with the speed circle. Adding multitasking did not affect movement fluidity, as shown by no significant difference in flying and idle percentages between tasks using the joystick (t(38) = -1.073, p = 0.290) or speed circle (t(38) = -0.268, p = 0.790).

\subsubsection*{\textbf{Path Length}}
See Fig.~\ref{fig:pathBoxPlot}. Path length averages were similar among tasks and Ring conditions, ranging from 376.67 to 408.86 meters. There are no significant differences in Ring tasks between joystick and speed circle techniques, with users traveling approximately 2.37\% less using the joystick ($t(38) = -0.676, p = 0.503$), consistent with the Magic Carpet study (see further below). No significant difference in path lengths was found in the target practice task ($t(38) = 1.303, p = 0.200$).

\subsubsection*{\textbf{Speed}}
The speed plot (Fig.~\ref{fig:speedBoxPlot}) shows that Rings+Joystick had a median speed of $3.5 m/s$ with high variability. Rings+Speed Circle shows a lower median speed of $2 m/s$ with slightly more consistency, probably because users did not always reach the maximum speed. Targets+Joystick has a median speed of $2.5 m/s$ with high variability, while Targets+Speed Circle has the highest median speed at $3.5 m/s$, also with high variability. The joystick technique supports higher speeds, especially in Rings tasks.

\subsubsection*{\textbf{Collisions}}
The collisions did not show significant differences (See Fig.~\ref{fig:collisionsBoxPlot}) between the techniques in the rings ($t(38) = 0.369, p = 0.714$) and the target practice ($t(38) = 0.186, p = 0.853$) tasks. More collisions occurred in the Target practice task than in the Rings task, indicating a slight decrease in performance with multitasking. This pattern holds for both joystick ($t(38) = -2.067, p = 0.045$) and speed circle ($t(38) = -2.569, p = 0.014$) techniques.

\subsection{Performance compared  to Magic Carpet}\label{sec:MCcompare}

\begin{figure*}[b]
    \centering
    \includegraphics[width=0.9\linewidth]{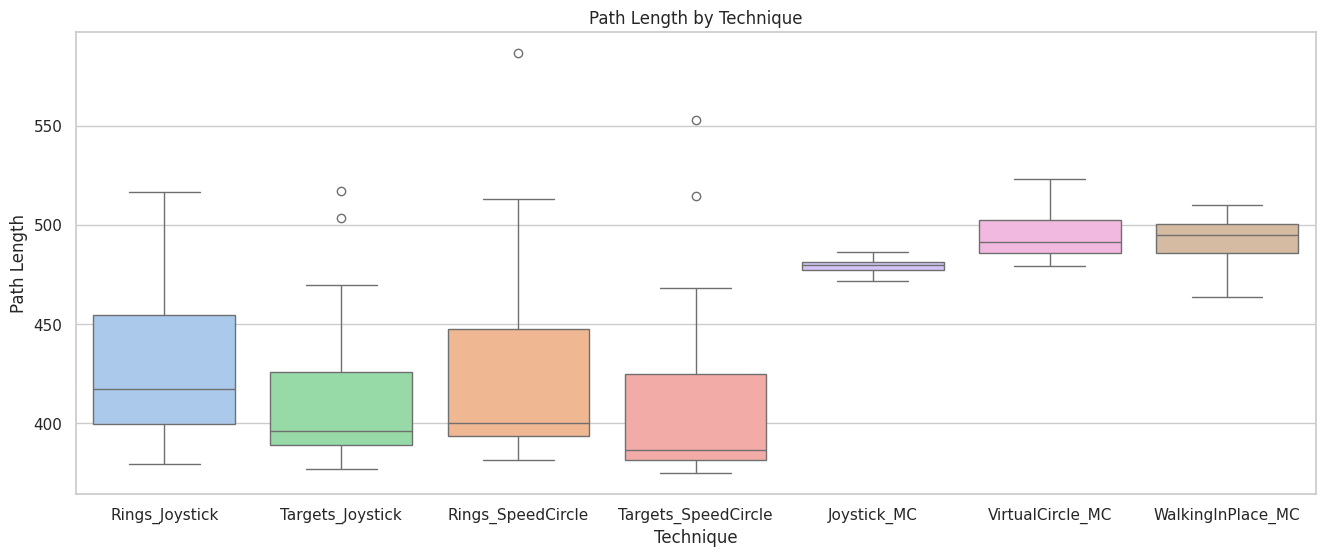}
    \caption{Comparison of the traveled path with each of our conditions and the Magic Carpet three conditions.}
    \label{fig:MCPath}
\end{figure*}

The joystick and speed circle techniques covered less distance than the Magic Carpet study (see Fig.~\ref{fig:MCPath}). This occurs even for the target task where a secondary subtask is active. Threading a shorter path indicates that the maneuverability of the technique is more efficient. Table~\ref{tab:pathMC} shows the significant differences after applying a Kruskal-Wallis H test and a Dunn test adjusted with Bonferroni.

\begin{table*}[t]
\small
  \centering
    \caption{Significance for the Path Length differences among conditions}
    \label{tab:pathMC}
    \resizebox{\dimexpr\textwidth-4pt\relax}{!}{
\begin{tabular}{l|lllllll}
                     & Rings Joy                     & Targets Joy                   & Rings SC                      & Targets SC                    & Joy MC                        & VC MC                         & WIP MC                        \\ \hline
Rings Joy      & 1.0000                         & 1.0000                         & 1.0000                         & 1.0000                         & 0.7339                         & \cellcolor[HTML]{34FF34}0.0017 & \cellcolor[HTML]{34FF34}0.0052 \\
Targets Joy    & 1.0000                         & 1.0000                         & 1.0000                         & 1.0000                         & \cellcolor[HTML]{34FF34}0.0300 & \cellcolor[HTML]{34FF34}0.0000 & \cellcolor[HTML]{34FF34}0.0000 \\
Rings SC   & 1.0000                         & 1.0000                         & 1.0000                         & 1.0000                         & 0.3482                         & \cellcolor[HTML]{34FF34}0.0004 & \cellcolor[HTML]{34FF34}0.0015 \\
Targets SC & 1.0000                         & 1.0000                         & 1.0000                         & 1.0000                         & \cellcolor[HTML]{34FF34}0.0072 & \cellcolor[HTML]{34FF34}0.0000 & \cellcolor[HTML]{34FF34}0.0000 \\
Joystick MC         & 0.7339                         & \cellcolor[HTML]{34FF34}0.0300 & 0.3482                         & \cellcolor[HTML]{34FF34}0.0072 & 1.0000                         & 1.0000                         & 1.0000                         \\
VC MC    & \cellcolor[HTML]{34FF34}0.0017 & \cellcolor[HTML]{34FF34}0.0000 & \cellcolor[HTML]{34FF34}0.0004 & \cellcolor[HTML]{34FF34}0.0000 & 1.0000                         & 1.0000                         & 1.0000                         \\
WIP MC   & \cellcolor[HTML]{34FF34}0.0052 & \cellcolor[HTML]{34FF34}0.0000 & \cellcolor[HTML]{34FF34}0.0015 & \cellcolor[HTML]{34FF34}0.0000 & 1.0000                         & 1.0000                         & 1.0000                        
\end{tabular}
}
\end{table*}

Although very few collisions occurred overall, there were significantly fewer collisions in the ring task with our technique, either using a joystick or speed circle for speed, than with the Magic Carpet conditions, except for Joystick, where the effect is seen but is insignificant. It is also visible that the target task distracts some participants enough so that they allow more collisions. Still, the distribution is more widespread, showing that some participants are unaffected, implying that training can improve their performance. Significance measures after applying a Kruskal-Wallis H test and a Dunn's Test adjusted with Bonferroni are in table~\ref{tab:collisionMC}. 


\begin{table*}[t]
  \centering
  \small
    \caption{Significance for the differences of collision events among conditions}
    \label{tab:collisionMC}
    \resizebox{\dimexpr\textwidth-12pt\relax}{!}{
\begin{tabular}{l|lllllll}
                     & \textbf{Rings Joy}            & \textbf{Targets Joy} & \textbf{Rings SC}             & \textbf{Targets SC} & \textbf{Joy MC} & \textbf{VC MC}                & \textbf{WIP MC}               \\ \hline
Rings Joy      & 1.0000                         & 0.2170                & 1.0000                         & 0.1254               & 1.0000           & \cellcolor[HTML]{34FF34}0.0101 & \cellcolor[HTML]{34FF34}0.0056 \\
Targets Joy    & 0.2170                         & 1.0000                & 0.1304                         & 1.0000               & 1.0000           & 1.0000                         & 1.0000                         \\
Rings SC   & 1.0000                         & 0.1304                & 1.0000                         & 0.0731               & 1.0000           & \cellcolor[HTML]{34FF34}0.0054 & \cellcolor[HTML]{34FF34}0.0029 \\
Targets SC & 0.1254                         & 1.0000                & 0.0731                         & 1.0000               & 1.0000           & 1.0000                         & 1.0000                         \\
Joy MC         & 1.0000                         & 1.0000                & 1.0000                         & 1.0000               & 1.0000           & 0.6829                         & 0.4678                         \\
VC MC    & \cellcolor[HTML]{34FF34}0.0101 & 1.0000                & \cellcolor[HTML]{34FF34}0.0054 & 1.0000               & 0.6829           & 1.0000                         & 1.0000                         \\
WIP MC   & \cellcolor[HTML]{34FF34}0.0056 & 1.0000                & \cellcolor[HTML]{34FF34}0.0029 & 1.0000               & 0.4678           & 1.0000                         & 1.0000                        
\end{tabular}
}
\end{table*}

\begin{figure*}[!b]
    \centering
    \includegraphics[width=0.9\linewidth]{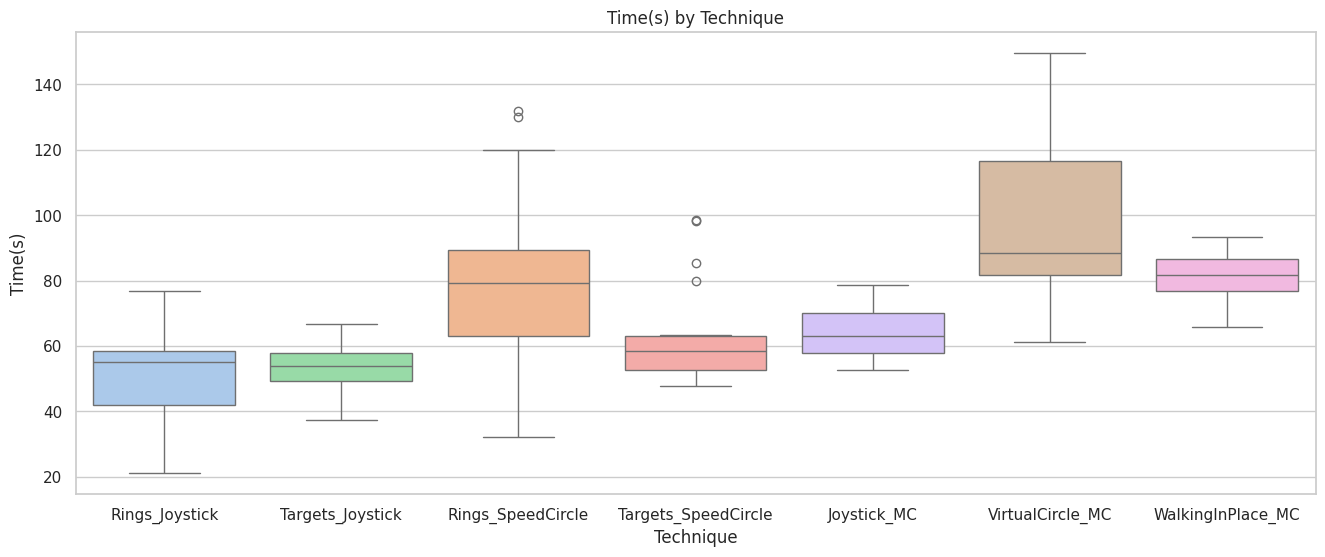}
    \caption{Comparison of the time to complete the task with each of our conditions and the Magic Carpet's three conditions.}
    \label{fig:MCTimes}
\end{figure*}

Finally, the results of the time performance show that our eye gaze steering technique, either with a joystick or a speed circle, is more efficient than the Magic Carpet conditions or at least similar, but not worse (cf. Fig~\ref{fig:MCTimes}). Although $Joystick\_MC$ is not significantly different from our joystick results (see Table~\ref{tab:timeMC} for significance), this is due to the lack of training with our technique, which causes the samples to spread. However, all participants in our experiment performed the rings before the target task. This induced a learning effect that considerably reduced variance. This also reduced the average time for the speed circle, which requires more learning than the joystick.

\begin{table*}[t]
  \centering
  \small
    \caption{Significance measures for the time to complete differences among conditions}
    \label{tab:timeMC}
    \resizebox{\dimexpr\textwidth-12pt\relax}{!}{
\begin{tabular}{l|lllllll}
                     & \textbf{Rings Joy}            & \textbf{Targets Joy}          & \textbf{Rings SC}             & \textbf{Targets SC}           & \textbf{Joy MC}               & \textbf{VC MC}                & \textbf{WIP MC}               \\ \hline
Rings Joy      & 1.0000                         & 1.0000                         & \cellcolor[HTML]{34FF34}0.0018 & 1.0000                         & 0.4893                         & \cellcolor[HTML]{34FF34}0.0000 & \cellcolor[HTML]{34FF34}0.0000 \\
Targets Joy    & 1.0000                         & 1.0000                         & \cellcolor[HTML]{34FF34}0.0010 & 1.0000                         & 0.3412                         & \cellcolor[HTML]{34FF34}0.0000 & \cellcolor[HTML]{34FF34}0.0000 \\
Rings SC   & \cellcolor[HTML]{34FF34}0.0018 & \cellcolor[HTML]{34FF34}0.0010 & 1.0000                         & 0.5095                         & 1.0000                         & 0.7758                         & 1.0000                         \\
Targets SC  & 1.0000                         & 1.0000                         & 0.5095                         & 1.0000                         & 1.0000                         & \cellcolor[HTML]{34FF34}0.0004 & \cellcolor[HTML]{34FF34}0.0256 \\
Joy MC         & 0.4893                         & 0.3412                         & 1.0000                         & 1.0000                         & 1.0000                         & \cellcolor[HTML]{34FF34}0.0114 & 0.2470                         \\
VC MC    & \cellcolor[HTML]{34FF34}0.0000 & \cellcolor[HTML]{34FF34}0.0000 & 0.7758                         & \cellcolor[HTML]{34FF34}0.0004 & \cellcolor[HTML]{34FF34}0.0114 & 1.0000                         & 1.0000                         \\
WIP MC   & \cellcolor[HTML]{34FF34}0.0000 & \cellcolor[HTML]{34FF34}0.0000 & 1.0000                         & \cellcolor[HTML]{34FF34}0.0256 & 0.2470                         & 1.0000                         & 1.0000                        
\end{tabular}
}
\end{table*}

\subsection{Discussion}

The usability testing of our VR steering system yielded highly promising results, as evidenced by the metrics from the SUS-presence, SSQ, and NASA-TLX questionnaires. The SUS scores indicated a strong presence, with an average score of $15.54$ out of 20 
$(77.7\%)$. This suggests that our system provides a high level of user immersion. The SSQ score, with an average total score of $21.25$ could be significant, especially when disorientation was scored higher, at $30.37$. However, the nausea factor is lower, $13.87$ on average. These symptoms are concerning but still lower than similar approaches like the Magic Carpet (with joystick SSQ=$39.52$; speed circle SSQ=$43.84$). This indicates that overall comfort was maintained.

The NASA-TLX total score of $28.91$, places the system within the medium workload category. Since multitasking was expected to overload the participant's work further, this result indicates that gaze-hand steering successfully accommodated multitasking.

There were significant differences in flying and idle times between techniques, with longer idle times occurring during both tasks when using the joystick compared to the speed circle. This difference was more pronounced in the Target task. The joystick caused users to stop moving more frequently to look around. This can be attributed to the ease of performing dual tasks with body gestures rather than physical input.



There are indications of slightly lower performance in reaching the rings using the speed circle. Regarding the Target task, the mean number of targets destroyed using the speed circle ($28.95$) is very similar to the joystick ($28.85$). This indicates that the technique does not affect the accuracy of the Target task.


Results compared with the Magic Carpet technique show that our technique allows for overall performance similar to the previous technique while allowing for a secondary simultaneous task. Some performance measures are better with our technique, such as collisions and distance traveled, arguably due to the speed circle and HMD setup improvements. 

Regarding body gestures, we argue that evaluating the user's position within the circle could be more accurately achieved by assessing the position of the torso rather than the head. Users reported difficulties looking upwards and attempting to navigate forward as their heads instinctively moved back. This resulted in a shorter distance from the circle's center, leading to slower velocities.

\section{Conclusion}
\label{sec:conclusion}

In this paper, we approached the problem of how a gaze-directed approach assisted by hand pointing could prevail relative to other known steering techniques. We designed a method that provides a free look, opening possibilities for multitasking during travel in virtual environments. 

We developed an application to study our technique and conducted a user study with 20 participants. Additionally, we analyzed our gaze-directed approach compared to the Magic Carpet to find any possible effects. The results help us understand how applying this method in eye-gaze-dominant applications affects exploration by decoupling gaze from directed steering while still being able to navigate simultaneously. 

One important finding is that performance does not decrease when introducing a new task while navigating. Users were able to complete objectives like passing through rings and shooting targets with success rates of 97.5\% and 96.3\%, respectively, which confirms the effectiveness of our gaze-directed technique. Based on these results, we argue that the system can be used to navigate virtual environments in a range of applications.

One limitation of our study is the collection of user cybersickness, presence, and workload inputs for each technique. The current results assess these variables in general rather than for specific techniques or tasks. A more detailed analysis could improve comparisons with the Magic Carpet and other studies in future work to provide clearer conclusions. An important feature to implement is to use the torso instead of the head position for speed control. We also suggest that hand-tracking instead of controllers be used in a new user study, as this would align with the current trend of gaze-dominant hands-free interaction in XR. 
\section*{Privacy/Ethics Statement}
This study, approved by an ethics board, examines a gaze-directed VR navigation technique for multitasking, using anonymized eye tracking data. Although our method protects privacy, broader applications should consider data sovereignty and informed consent to mitigate unintended profiling or misuse risks.

\bibliography{sample-base}

\end{document}